\documentclass[12pt,titlepage]{article}
\usepackage{graphicx}
\newcommand{\be}{\begin{equation}}
\newcommand{\ee}{\end{equation}}

\newcommand{\magsim}
{\ \lower2pt\hbox{$\sim $}\mkern-14mu \raise2pt\hbox{$>$}\ }
\newcommand{\br}{{\bf r}}
\newcommand{\brd}{\dot{\bf r}}
\newcommand{\yr}{{\rm \, yr}}
\newcommand{\cm}{{\rm \, cm}}
\newcommand{\gm}{{\rm \, g}}
\newcommand{\km}{{\rm \, km}}

\newcommand{\s}{{\rm \, s}}
\def\refnew#1{(\ref{#1})}
\begin{document}

\begin{center}
{\tiny .  }
\vskip 0.75truein
{\LARGE \bf Chaotic Motions of F-Ring Shepherds}
\vskip 0.75truein
\baselineskip16pt
{\Large \bf Peter Goldreich} \\
\vskip 0.15truein
{\large \it California Institute of Technology} \\
{\large \it Pasadena CA 91125} \\
{\large \it E-mail: pmg@gps.caltech.edu}
\vskip 0.25truein
{\large and}
\vskip 0.25truein
{\Large \bf Nicole Rappaport} \\
\vskip 0.15truein
{\large \it Jet Propulsion Laboratory} \\
{\large \it California Institute of Technology}  \\
{\large \it Pasadena CA 91109} \\
{\large \it E-mail: Nicole.J.Rappaport@jpl.nasa.gov} \\
\large
\vskip 0.75truein
Submitted to Icarus on May 14, 2002 \\
\vskip 0.75truein
\baselineskip20pt
Number of pages: 23 \\
Number of tables: 1 \\
Number of figures: 12

\end{center}

\pagebreak

\noindent {\bf Proposed Running Head:} Chaotic Motions of F-Ring Shepherds

\vskip 0.5truein
\noindent {\bf Editorial correspondence to:} \\
\baselineskip14pt
\noindent Professor Peter Goldreich \\
\noindent California Institute of Technology \\
\noindent MS 170-25 \\
\noindent Pasadena CA 91125 \\
\noindent Phone: (626) 395 6193 \\
\noindent Fax: (626) 585 1917 \\
\noindent E-mail: pmg@gps.caltech.edu
\baselineskip29pt
\pagebreak

%Begin summary template
\begin{center}
{\bf ABSTRACT}
\end{center}

\begin{minipage}{5.5in}
\noindent
\small

Recent HST images of the Saturnian satellites Prometheus and Pandora
show that their longitudes deviate from predictions of ephemerides
based on Voyager images. Currently Prometheus is lagging and Pandora
leading these predictions by somewhat more than $20^\circ$. We show
that these discrepancies are fully accounted for by gravitational
interactions between the two satellites. These peak every $24.8\,{\rm
d}$ at conjunctions and excite chaotic perturbations. The Lyapunov
exponent for the Prometheus-Pandora system is of order $0.35\yr^{-1}$
for satellite masses based on a nominal density of $1.3 \gm\cm^{-3}$.
Interactions are strongest when the orbits come closest together. This
happens at intervals of $6.2\yr$ when their apses are anti-aligned.
In this context we note the sudden changes of opposite signs in the 
mean motions of Prometheus and Pandora at the end of 2000 occured shortly 
after their apsidal lines were anti-aligned.

\vskip 0.25truein


\normalsize
\end{minipage}
%End summary template

\pagebreak

\section{INTRODUCTION}

Orbits for Pandora and Prometheus in the form of precessing ellipses
of fixed shape were fit to Voyager data by Synnott {\it et al.} (1981,
1983) and Jacobson (personal communication).  Mean motions were
determined from images and precession rates were calculated to be
consistent with the gravity field of the Saturnian system (Nicholson
and Porco 1988; Campbell and Anderson 1989).

Observations with HST made during the 1995-1996 Sun and Earth ring
plane crossings led to the discovery that Prometheus was lagging its
predicted longitude based on the Voyager ephemeris by about
$20^{\circ}$ (Bosh and Rivkin 1996, Nicholson {\it et al.}
1996). Subsequently McGhee (2000) found that Pandora was leading the
Voyager ephemeris prediction by a similar amount. These discrepancies
have been confirmed by French {\it et al.} (1999, 2000, 2001, 2002),
Murray {\it et al.} (2000), McGhee {\it et al.} (2001), and Evans
(2001).

These and other researchers looked for a dynamical origin of the
longitude discrepancies.  Several hypotheses were investigated
including: perturbations exerted by an undetected coorbital satellite
of Prometheus (see French {\it et al.} 1998); interactions with clumps
in the F ring, or 1 to 5 km objects in the F ring, or the F ring
itself (Showalter {\it et al.} 1999a, 1999b); long-term resonance
dynamics (Dones {\it et al.} 1999), and chaos (Dones {\it et al.}
2001). However, none of these attempts provide a clear resolution of
this puzzle. That is the goal of our paper.

We focus on direct interactions between Pandora and Prometheus because
their longitude discrepancies have comparable magnitudes and opposite
signs (French {\it et al.} 2002). This suggests that the satellites
are exchanging angular momentum and energy and that their orbits are
chaotic. Results from orbit integrations presented in \S 2 of the
paper confirm this suspicion.  Implications of these findings for
estimates of the age of Saturn's rings are discussed in \S 3. This section
also reproduces evidence from French {\it et al.} (2002) that supports
our finding that sudden changes in mean motion tend to occur around times when
the satellites' apses are anti-aligned.  

\section{Confirmation of Chaos}

\subsection{Calculational Method}

Working in a planet centered coordinate system and adopting conventional
notation, the vector equation of motion for each satellite reads

\begin{eqnarray}
{d^2\br_i\over dt^2}=& - &{GM\br_i\over r_i^3}\left[\left (1+{m_i\over
M}\right){3\over 2} J_2\left({R\over r_i}\right)^2 -{15\over
8}J_4\left({R\over r_i}\right)^4 + {35\over 16}J_6\left({R\over
r_i}\right)^6\right]\\
& - & Gm_j\left({\br_i-\br_j\over
\left|\br_i-\br_j\right|^3} + {\br_j\over r_j^3}\right) \nonumber ,
\label{eq:motion}
\end{eqnarray}

where $i$ and $j$ ($i\neq j$) assume values 1 and 2. We integrate only
four first order scalar differential equations for each body since to
observational accuracy the orbits of Prometheus and Pandora lie in
Saturn's equatorial plane.

Equations \refnew{eq:motion} admit energy and angular momentum
integrals given by

\begin{eqnarray}
E&=&{1\over 2}\left[m_1\left|\brd_1\right|^2 + m_2\left|\brd_2\right|^2
-{\left|m_1\brd_1 + m_2\brd_2\right|^2\over M+m_1+m_2}\right] - {Gm_1m_2\over
\left|\br_1-\br_2\right|} \\
&-&{GMm_1\over r_1}\left[1 - J_2\left({R\over r_1}\right)^2 - J_4\left({R\over
r_1}\right)^4 - J_6\left({R\over r_1}\right)^6\right]  \nonumber \\
&-&{GMm_2\over r_2}\left[1 - J_2\left({R\over r_2}\right)^2 - J_4\left({R\over
r_2}\right)^4 - J_6\left({R\over r_2}\right)^6\right] \nonumber ,
\label{eq:E}
\end{eqnarray}
and
\be
H=m_1\left(\br_1\times\brd_1\right)+m_2\left(\br_2\times\brd_2\right)
-{\left(m_1\br_1+m_2\br_2\right)\times\left(m_1\brd_1+m_2\brd_2\right)\over
M+m_1+m_2}.
\label{eq:H}
\ee

Numerical integrations of the equations of motion are carried out
using the algorithm of Bulirsch and Stoer (1980) which offers the
luxury of a variable time step. Fractional changes in total energy and
angular momentum are of order $10^{-10}$ for integrations of
$10^3\yr$.  For comparison, jumps in these quantities are of order
$10^{-5}$ at each conjunction.

Initial conditions are computed from Jacobson's equinoctial elements
(Jacobson, personal communication) and the transformation between
cylindrical elements and epicyclic elements derived in
Borderies-Rappaport and Longaretti (1994).  The same transformation is
applied to compute the epicyclic eccentricity and the epicyclic mean
longitude at each output step. We also output values of the angular
momentum and energy (neglecting the interaction term) for each
satellite, and the total angular momentum and energy (including the
interaction term).

To compute Lyapunov exponents we integrate the orbits of two shadow
bodies whose initial conditions differ slightly from those of
Prometheus and Pandora.  We reset the state vector of the shadow
bodies to reduce the magnitude of their phase space separation from
the physical bodies whenever it exceeds a preset tolerance. In
practice rescaling is done when the longitudinal separations that
dominate the phase space separations are slightly less than $10^{-4}$
radians.

\subsection{Results}

Variations over 20 years of orbital longitudes for Prometheus and
Pandora are displayed in Figs. 1 and 2. Initial values for the
satellites' epicyclic eccentricities and mean motions obtained
following the prescription described in \S 2.1 are presented in Table
1; this table also contains the ratios of the satellites' masses to
Saturn's mass. The simulation begins with Prometheus at periapse and
Pandora at apoapse and with the satellites' apsidal lines in phase.
To emphasize the chaotic irregularities of the mean motions, we
subtract a drift rate based on the initial mean motion from the
longitude of each satellite.  These figures reproduce the
characteristics of the puzzling longitude discrepancies reported in
the papers referenced in
\S 1.  Line widths are due to epicyclic longitude oscillations which
have full amplitudes of $4e$ radians.

\underbar{Table I}. Initial values of Prometheus and Pandora
eccentricities and mean motion, and masses scaled to Saturn's mass.

\vskip 0.25truein

\begin{tabular}{|l|c|c|c|}
\hline
Satellite  & $e$                   & $n$ (rd/s)
& $m_i/M$ \\
\hline
Prometheus & $2.29 \times 10^{-3}$ & $1.1864 \times 10^{-4}$
& $1.19 \times 10^{-9}$ \\
Pandora    & $4.37 \times 10^{-3}$ & $1.1571 \times 10^{-4}$
& $7.65 \times 10^{-10}$ \\
\hline
\end{tabular}

\vskip 0.5truein

Figure 3 displays the difference between the apsidal angles over 20
years, and Fig. 4 shows the behavior of the magnitude of the relative
eccentricity vector.  The two orbits come closest together and the
magnitude of the relative eccentricity peaks when the apsidal lines
are anti-aligned.  This occurs at about $t = 3.1\yr$, $9.3\yr$, and
$15.5\yr$.  Comparison with Figs. 1 and 2 reveals that these mark the
times at which abrupt changes in the satellites' mean motions take
place. Note that net changes in mean motion around $t = 3.1\yr$ and
$15.5\yr$ years are of comparable magnitude but opposite sign, but
that the net change in mean motion that takes place around $t=9.3\yr$
is much smaller. This is another indication of chaos.

Additional evidence for chaos is found in plots of epicyclic
eccentricity vs. time shown in Figs. 5 and 6.  Two distinct types of
eccentricity variation are apparent.  Small jumps occur at
conjunctions separated by about 24.8 days. These have magnitudes of
order $\mu \left( a / \Delta r \right)^2 \sim 10^{-5}$, where $\mu$ is
the mass of the perturbing satellite divided by the mass of Saturn,
$a$ is the mean orbit radius, and $\Delta r$ is the radial distance
between the satellites at conjunction.  As expected, the largest jumps
occur when the satellites' apses are near anti-alignment.
Quasi-periodic variations of eccentricity are associated with the
relative apsidal precession period of 6.2 years.  They arise from
secular perturbations which promote the exchange of angular momentum
but not of energy. Since secular eccentricity variations are entirely
due to angular momentum exchanges, Prometheus's and Pandora's are 180
degrees out of phase.  Although the secular variations are somewhat
larger than the jumps, they are small in comparison to the mean
eccentricity.  Their small size is a consequence of the dominance of
Saturn's oblate gravitational equipotentials in forcing the
differential precession; secular terms in the satellites' interaction
potential contribute only a small fraction of the differential
precession rate.  Eccentricity jumps and secular eccentricity
variations are not the entire story, nor even the most important part
of it.  That distinction goes to the lack of periodicity over the
differential precession cycle which is a clear signature of chaos.

To prove that the mean motion variations arise from chaos, we compute
the Lyapunov exponent for the Prometheus-Pandora system. Figure 7
illustrates its behavior over an interval of 3000 years. The figure
also includes a dashed line showing a constant plus $\left(
\log t \right) / t$ fit to the final point of the solid curve. This is
the behavior that would be expected in the absence of chaos. Evidence
for chaos is overwhelming. The Lyapunov exponent is of order $0.35\yr^{-1}$.

Figures 8-11 derived from data spaced by $0.1{\rm \, d}$ show
perturbations near conjunctions in greater detail. Variations of
epicyclic eccentricity are depicted in Figs. 8 and 9 while Figs. 10
and 11 provide data on energy and angular momentum accrued during the
numerical integration. Each panel covers an interval of one year
centered either on $t = 3.1\yr$, when the apses are anti-aligned, or
on $t = 6.1\yr$, when they are aligned. Perturbations are noticeably
larger during the former than during the latter. Fractional jumps of
the energy and angular momentum of each satellite at conjunctions are
$\sim \mu \left( a / \Delta r \right)^2$ and $\sim \mu \Delta e \left(
a /\Delta r \right)^3$, respectively.\footnote{$\Delta e$ is the
magnitude of the relative eccentricity vector.} Each of these is of
order $10^{-5}$. That our estimates are reasonable can be seen by
noting that the energy and angular momenta of Prometheus and Pandora
are $\sim 10^{33}\gm\cm^2\s^{-2}$ and $\sim 10^{37}\gm
\cm^2\s^{-1}$, and that their jumps are $\sim 10^{27}\gm
\cm^2\s^{-2}$ and $\sim 10^{31}\gm\cm^2\s^{-1}$. Spikes seen in the
plots of energy and angular momentum are arise from the strong
interactions near conjunctions. Their widths of a few hours are
marginally resolved.

\section{Discussion}

The suggestion that interactions between Saturn's F-ring shepherds
make their motions chaotic is not new. It was raised long ago in an
article we wrote with Scott Tremaine (Borderies {\it et al.}
1984). For us its confirmation is almost like a dream come true. Below
we quote from this earlier work because a full 20 years after it was
written we could scarcely improve upon it.\footnote{Our article was
prepared for a conference held in the summer of 1982.} We have,
however, added a footnote to introduce modern notation.

``As external satellites extract angular momentum from the rings,
their orbits expand.  Calculations based on the formula for the linear
satellite torque predict remarkably short time scales for the
recession of close satellites from the rings (Goldreich and Tremaine
1982).'' ``Thus these short time scales remain perhaps the most
intriguing puzzle in planetary ring dynamics.'' ``The severe nature of
the problem is well-illustrated by the system composed of the F-Ring
and its two shepherd satellites, 1980S26 and 1980S27, hereafter called
S26 and S27.''\footnote{1980S26 and 1980S27 were renamed Pandora and
Prometheus by the International Astronomical Union.} ``The outward
movement of the system could be reduced if S26 were involved in a
resonance with a more massive outer satellite.'' ``No resonance has
been found linking S26 with one or more outer satellites.  What might
this imply?  Are the F ring and its shepherd satellites very young?
Does the long sought resonance exist awaiting detection?  A most
interesting possibility is that S26 is transferring angular momentum
to Mimas even though the two bodies are not in an exact orbital
resonance.  This could be accomplished if the motion of S26 were
chaotic, i.e., if the value of its mean motion were undergoing a slow
random walk.  We have proven that if the mean longitude of S26 were
subject to a significant random drift, in addition to its dominant
secular increase, angular momentum transfer to Mimas would take place
by virtue of the near resonance between S26 and Mimas.  By significant
drift, we mean of order one radian on the circulation time scale of
the critical argument associated with the near resonance.  To check
this hypothesis, we must first determine whether the orbital motion of
S26 is chaotic.  To do so we need to investigate the perturbations of
its orbit produced by S27.  Solution of this and the other outstanding
theoretical problems will await results of ongoing research.''

Proving that the F-ring shepherds move chaotically as the result of
their mutual interactions is an important step. What it implies about
the lifetime of Saturn's rings remains to be determined.

We would be remiss if we failed to mention that chaotic motions of
Prometheus and Pandora are discussed in Poulet and Sicardy (2001),
Dones {\it et al.}  (2001), and French {\it et al.} (2002). Poulet and
Sicardy (2001) investigate the long term evolution of the system and
find intervals of chaos. Dones {\it et al.} (2001) suggest that chaos
might account for unexplained motions of the satellites but do not
identify the specific mechanism responsible for creating it. French
{\it et al} (2002) raise the possibility that changes of opposite sign
in the mean motions of Prometheus and Pandora may signal the exchange
of energy between their orbits. However, they do not simulate the
effects of interactions between the shepherds. Instead they present
evidence that external satellites can excite chaotic motions of test
particles in a portion of a region of $2\times 10^3\km$ width covering
the semimajor axes of both shepherds.

We close this paper by displaying evidence in support of our finding that
abrupt changes in mean motions tend to occur around times during which the
satellites' apses are anti-aligned. Figure 12 reproduces, with embelishments,  
panels from Figs. 3 and 5 of French {\it et al.} (2000). It shows that
Prometheus and Pandora underwent oppositely directly changes of their mean
motions around the end of year 2000, shortly after the time at which their 
apsidal longitudes differed by $180^\circ$. 

\section{Acknowledgments}

We thank R. French for providing us with a copy his preprint which
contains the latest results on the motions of Prometheus and Pandora, and
for allowing us to reproduce from it the material in our Fig. 12. 
We are grateful to R. Jacobson for advice on initializing our integrations in a
manner compatible with the Voyager ephemerides. Research by PG was
supported by NSF grant AST-0098301 and that by NR by NASA Planetary
Geology and Geophysics grant 344-30-53-02.

\section{References}

\baselineskip16pt

\qquad Borderies, N., P. Goldreich, and S. Tremaine 1984: Unsolved
Problems in Planetary Ring Dynamics, in \underbar{Planetary Rings},
Editors: Richard Greenberg and Andr\'e Brahic, The University of
Arizona Press, Tucson, Arizona, pages 713-734.

\vskip 0.15truein

Borderies-Rappaport N., and P.-Y. Longaretti 1994: Test Particle
Motion around an Oblate Planet, {\it Icarus} {\bf 107}, 129-141.

\vskip 0.15truein

Bosh, A.S. and A.S. Rivkin 1996: Observations of Saturn's inner
satellites during the May 1995 ring-plane crossing, {\it Science} {\bf
272}, 518-521.

\vskip 0.15truein

Bulirsch, R., and J. Stoer 1980: \underbar{Introduction to Numerical
Analysis}, Springer Verlag, New York.

\vskip 0.15truein

Campbell, J.K. and J.D. Anderson 1989: Gravity Field of the Saturnian
System from {\sl PIONEER} and {\sl VOYAGER} Tracking Data, {\it
Astron. J.} {\bf 97}, 1485-1495.

\vskip 0.15truein

Dones, L., M.R. Showalter, R.G. French, and J.J.  Lissauer 1999: The
Perils of Pandora, {\it BAAS} DPS Meeting \#31.

\vskip 0.15truein

Dones, L., H.F. Levison, J.J. Lissauer, R.G. French, and C.A. Mcghee
2001: Saturn's Coupled Companions, Prometheus and Pandora, {\it BAAS}
DPS Meeting \#33.

\vskip 0.15truein

Evans, M. 2001: \underbar{The determination of orbits from spacecraft
imaging}, Ph. D. Dissertation, Queen Mary College, University of
London.

\vskip 0.15truein

French, R.G., K.J. Hall, C.A. Mcghee, P.D.  Nicholson, J. Cuzzi,
L. Dones, and J. Lissauer 1998: The Perigrinations of Prometheus, {\it
BAAS} DPS Meeting \#30.

\vskip 0.15truein

French, R.G., C.A. Mcghee, P.D. Nicholson, L. Dones, and J. Lissauer
1999: Saturn's wayward shepherds: Pandora and Prometheus, {\it BAAS}
DPS Meeting \#31.

\vskip 0.15truein

French, R.G., C. Mcghee, L. Dones, and J.J. Lissauer 2000: The
Peripatetics of Prometheus and Pandora, {\it BAAS} DPS Meeting \#32.

\vskip 0.15truein

French, R.G., C.A., Mcghee, L. Dones, and J.J.  Lissauer 2001: The
Renegade Roamings of Prometheus and Pandora, {\it BAAS} DPS Meeting
\#33.

\vskip 0.15truein

French, R.G., C.A. Mcghee, L. Dones, and J.J. Lissauer 2002: Saturn's
Wayward Shepherds: The Perigrinations of Prometheus and Pandora, {\it
Icarus}, submitted for publication.

\vskip 0.15truein

Goldreich, P. and S. Tremaine 1982: The Dynamics of Planetary Rings,
{\it Ann. Rev. Astron. Astrophys.} {\bf 20}, 249-283.

\vskip 0.15truein

Mcghee, C. A. 2000: \underbar{Comet Shoemaker-Levy's 1994 collision
with Jupiter and Saturn's} \underbar{1995 ring plane crossing},
Ph. D. Dissertation, Cornell University, Ithaca, NY.

\vskip 0.15truein

Mcghee, C.A., P.D. Nicholson, R.G. French, and K.J. Hall 2001: HST
Observations of Saturnian Satellites during the 1995 Ring Plane
Crossings, {\it Icarus} {\bf 152}, 282-315.

\vskip 0.15truein

Murray, C.D., M.W. Evans, C.C. Porco, and M.R.  Showalter 2000: The
Orbits of Prometheus, Pandora and Atlas in 1980 and 1981, {\it BAAS}
DPS Meeting \#32.

\vskip 0.15truein

Nicholson, P.D., and C.C. Porco 1988: A New Constraint on Saturn's
Zonal Gravity Harmonics from Voyager Observations of an Eccentric
Ringlet, {\it J. Geophys. Res.} {\bf 93}, 10,209-10,224.

\vskip 0.15truein

Nicholson, P.D., M.R. Showalter, L. Dones, R.G. French, S.M. Larson,
J.J. Lissauer, C.A. Mcghee, P. Seitzer, B. Sicardy, and G.E. Danielson
1996: Observations of Saturn's ring--plane crossings in August and
November 1995, {\it Science} {\bf 272}, 509-515.

\vskip 0.15truein

Poulet, F. and B. Sicardy 2001: Dynamical evolution of the
Prometheus-Pandora System,{\it Mon. Not. R. Astron. Soc.} {\bf 322},
343-355.

\vskip 0.15truein

Showalter, M.R., L. Dones, and J.J.  Lissauer 1999a: Interactions
between Prometheus and the F Ring, {\it BAAS} DPS Meeting
\# 31.

\vskip 0.15truein

Showalter, M.R., L. Dones, and J.J.  Lissauer 1999b: Revenge of the
Sheep: Effects of Saturn's F Ring on the Orbit of Prometheus, {\it
BAAS} DPS Meeting \# 31.

\vskip 0.15truein

Synnott, S.P., C.F. Peters, B.A. Smith, and L.A. Morabito 1981: Orbits
of the Small Satellites of Saturn, {\it Science} {\bf 212}, 191-192.

\vskip 0.15truein

Synnott, S.P., R.J. Terrile, R.A. Jacobson, and B.A. Smith 1983:
Orbits of Saturn's F Ring and Its Shepherding Satellites, {\it Icarus}
{\bf 53}, 156-158.

\vfil\eject

\section{Figure Captions}

\vskip0.5truecm

\noindent\underbar{FIGURE 1}: Prometheus longitude from numerical
integration as a function of time. A drift rate based on the initial
mean motion is subtracted from the longitude. Units are degrees and
years.

\vskip0.5truecm

\noindent\underbar{FIGURE 2}: Pandora longitude from numerical
integration as a function of time.  A drift rate based on the initial
mean motion is subtracted from the longitude. Units are degrees and
years.

\vskip0.5truecm

\noindent\underbar{FIGURE 3}: Difference between the epicyclic
apsidal longitudes (in degrees) of Prometheus and Pandora over 20
years.

\vskip0.5truecm

\noindent\underbar{FIGURE 4}: Magnitude of the Prometheus and Pandora
relative eccentricity vector. Note that the peaks correspond to
anti-aligned apses.

\vskip0.5truecm

\noindent\underbar{FIGURE 5}: Prometheus epicyclic eccentricity
as a function of time.  The epicyclic frequency is computed from the
state in rectangular coordinates following Borderies-Rappaport and
Longaretti (1994).

\vskip0.5truecm

\noindent\underbar{FIGURE 6}: Pandora epicyclic eccentricity
as a function of time.  The epicyclic frequency was computed from the
state in rectangular coordinates following Borderies-Rappaport and Longaretti
(1994).

\vskip0.5truecm

\noindent\underbar{FIGURE 7}: Lyapunov exponent for the Prometheus-Pandora
system over a period of $3\times 10^3\yr$ (solid line). The dashed
line depicts a constant $+ \left( \log t \right) / t$ fit to the final
point of the solid curve. The unit for the Lyapunov exponent is
$\yr^{-1}$.

\vskip0.5truecm

\noindent\underbar{FIGURE 8}: Prometheus and Pandora epicyclic
eccentricities during an interval of one year centered on $t=3.1\yr$
when the apses are anti-aligned.

\vskip0.5truecm

\noindent\underbar{FIGURE 9}: Prometheus and Pandora epicyclic
eccentricities during an interval of one year centered on $t=6.2\yr$
when the apses are aligned.

\vskip0.5truecm

\noindent\underbar{FIGURE 10}: Prometheus (solid lines) and Pandora
(dotted lines) variations in energy (in $\gm\cm^2\s^{-2}$) and angular
momentum (in $\gm\cm^2\s^{-1}$) during an interval of one year
centered on $t=3.1\yr$ when the apses are anti-aligned. The dot-dashed
lines display differences between current and initial total energies
and angular momenta.

\vskip0.5truecm

\noindent\underbar{FIGURE 11}: Prometheus (solid lines) and Pandora
(dotted lines) variations in energy (in $\gm\cm^2\s^{-2}$) and angular
momentum (in $\gm\cm^2\s^{-1}$) during an interval of one year
centered on $t=6.2\yr$ when the apses are aligned. The dot-dashed
lines display differences between current and initial total energies
and angular momenta.

\vskip0.5truecm

\noindent\underbar{FIGURE 12}: Evidence for sudden jumps in the mean motions
of Prometheus and Pandora at the end of year 2000. Reproduced, with permission, 
from French {\it et al.} (2002). The solid vertical lines mark the time at
which the satellites' apses were anti-aligned.

\includegraphics{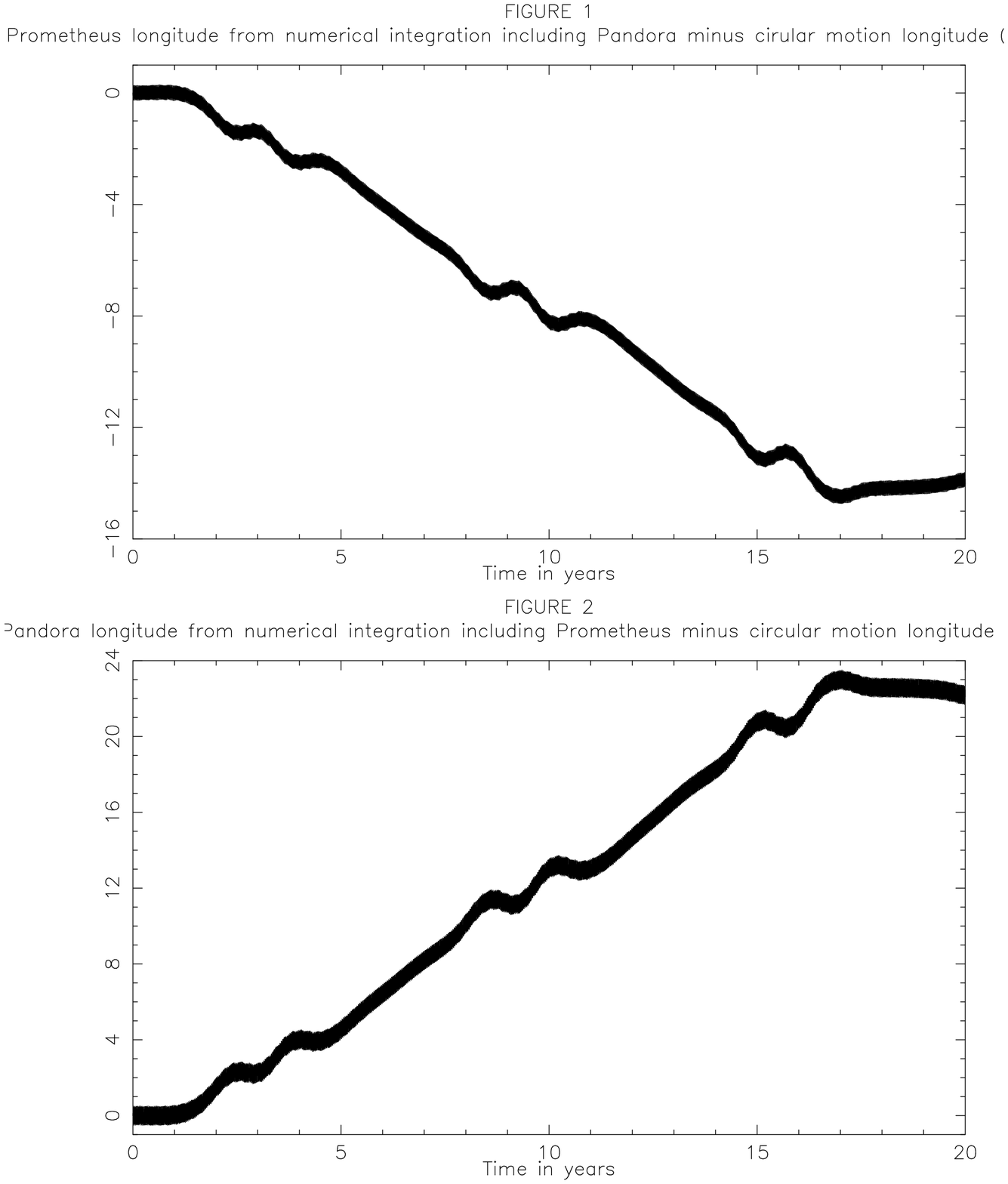}
\includegraphics{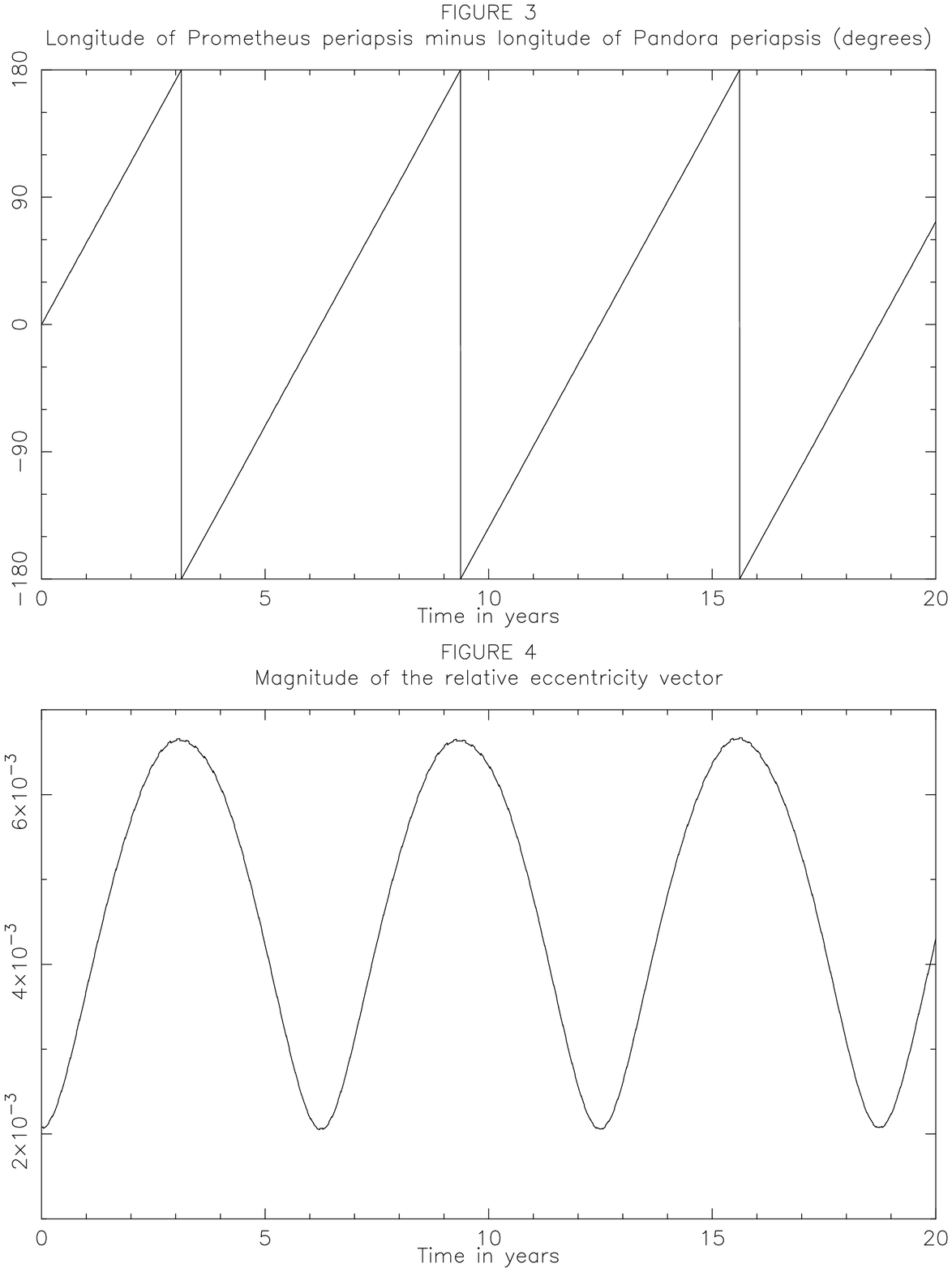}
\includegraphics{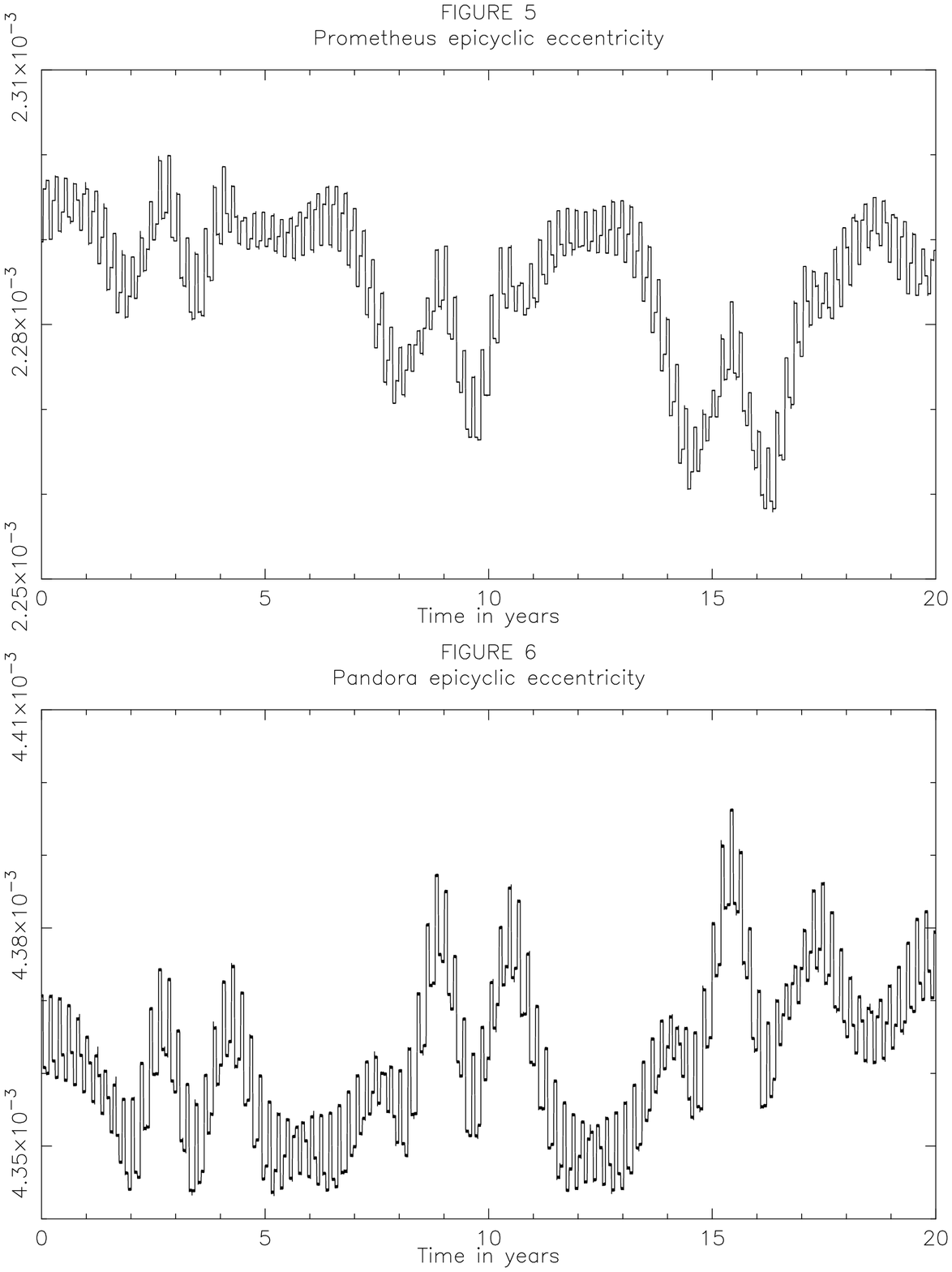}
\includegraphics{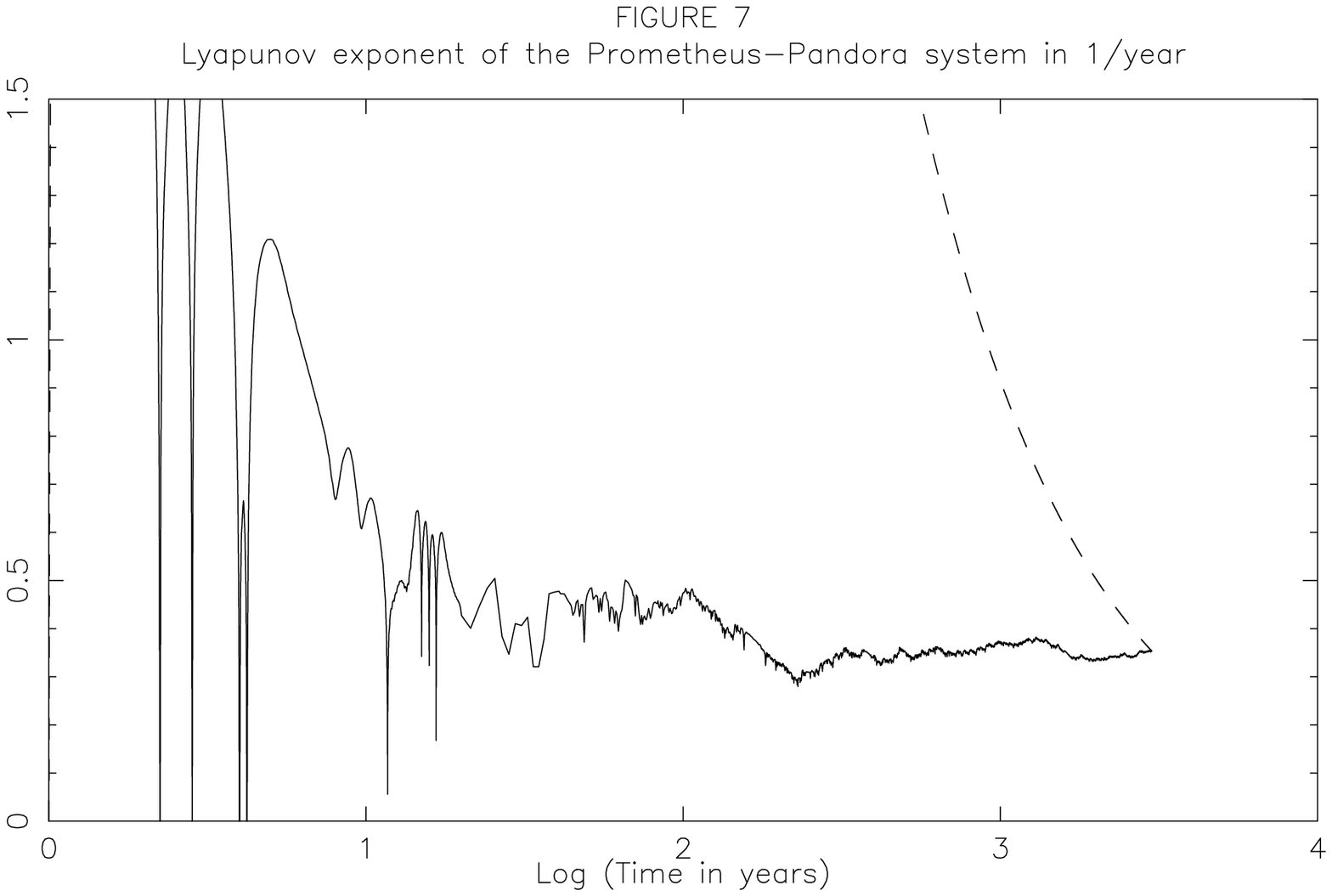}
\includegraphics{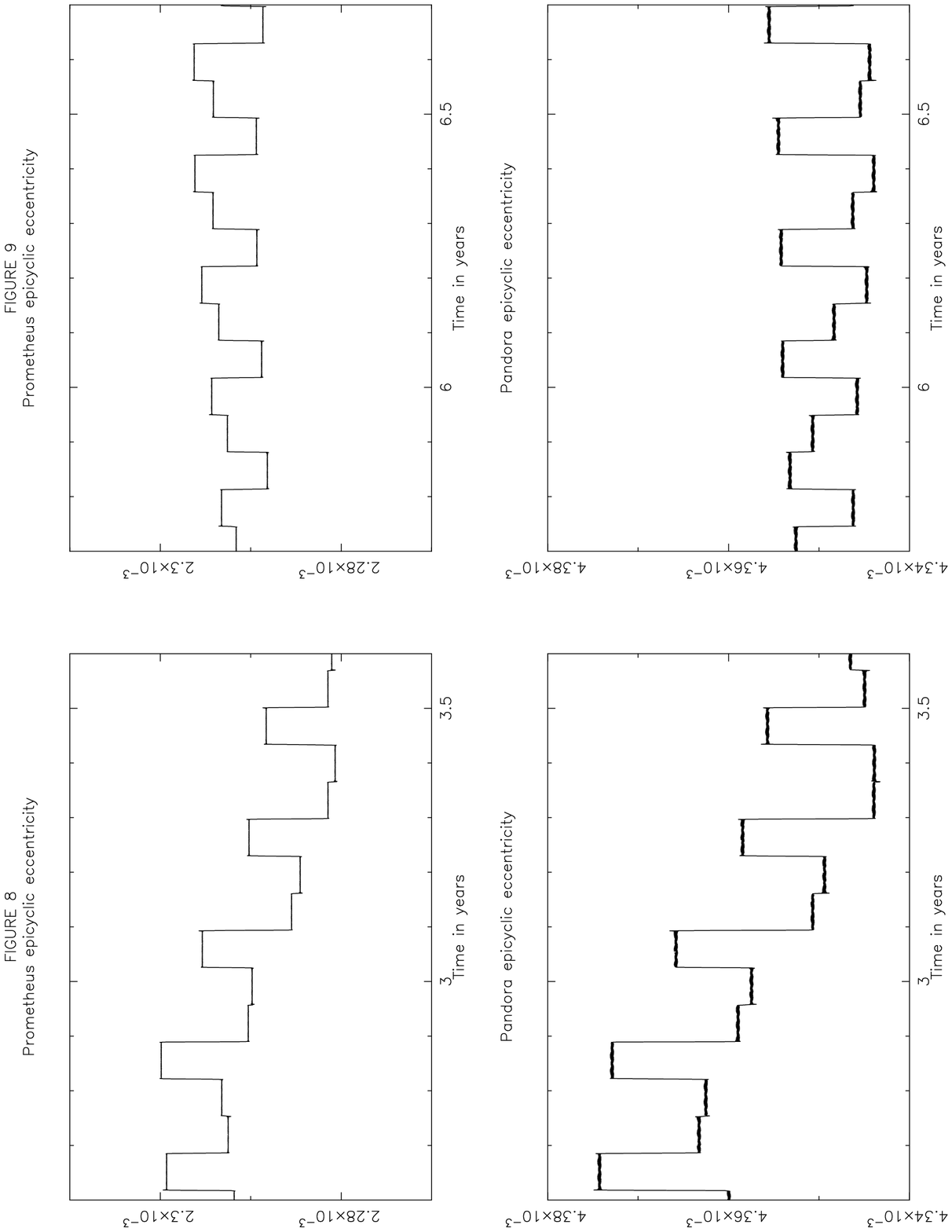}
\includegraphics{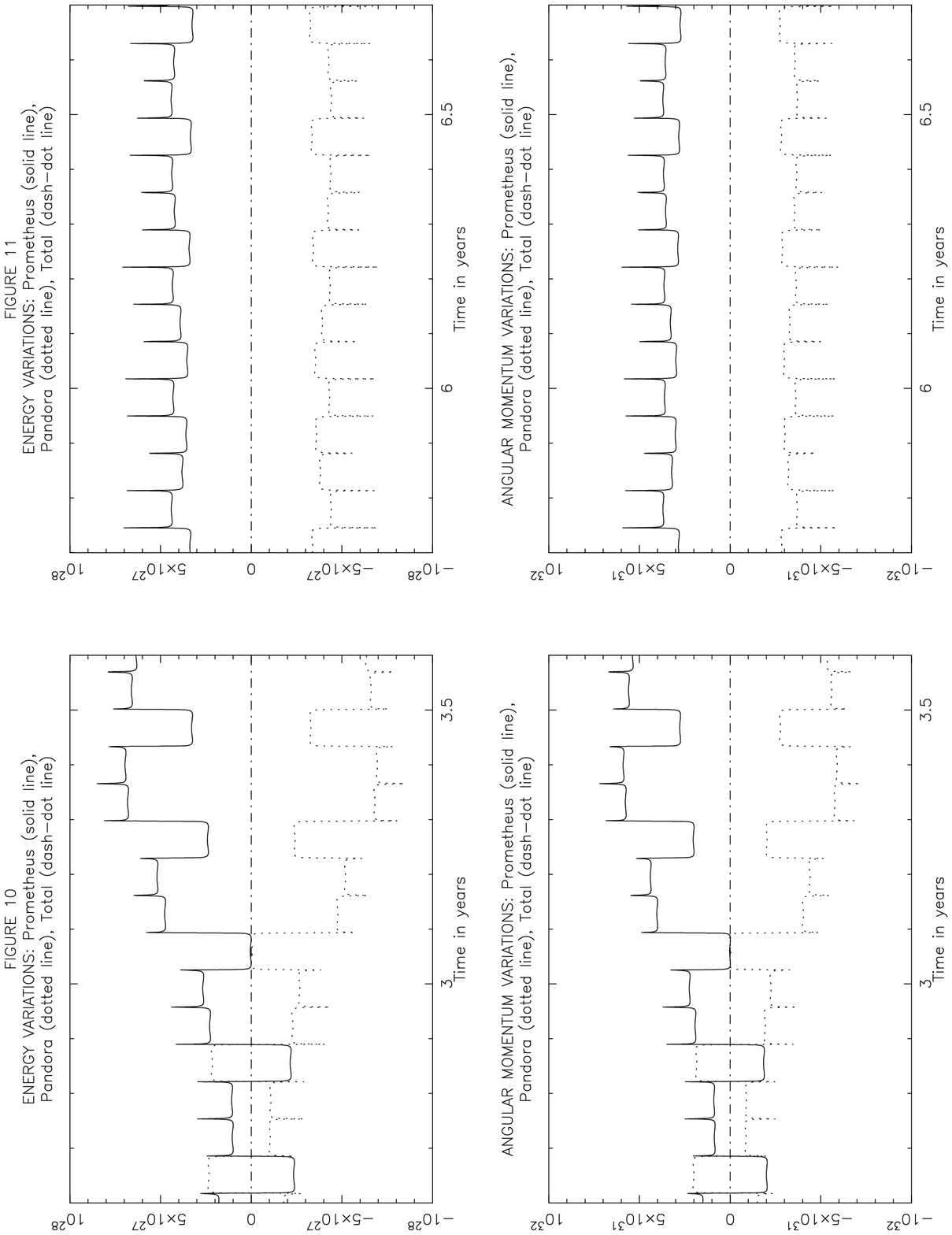}
\includegraphics{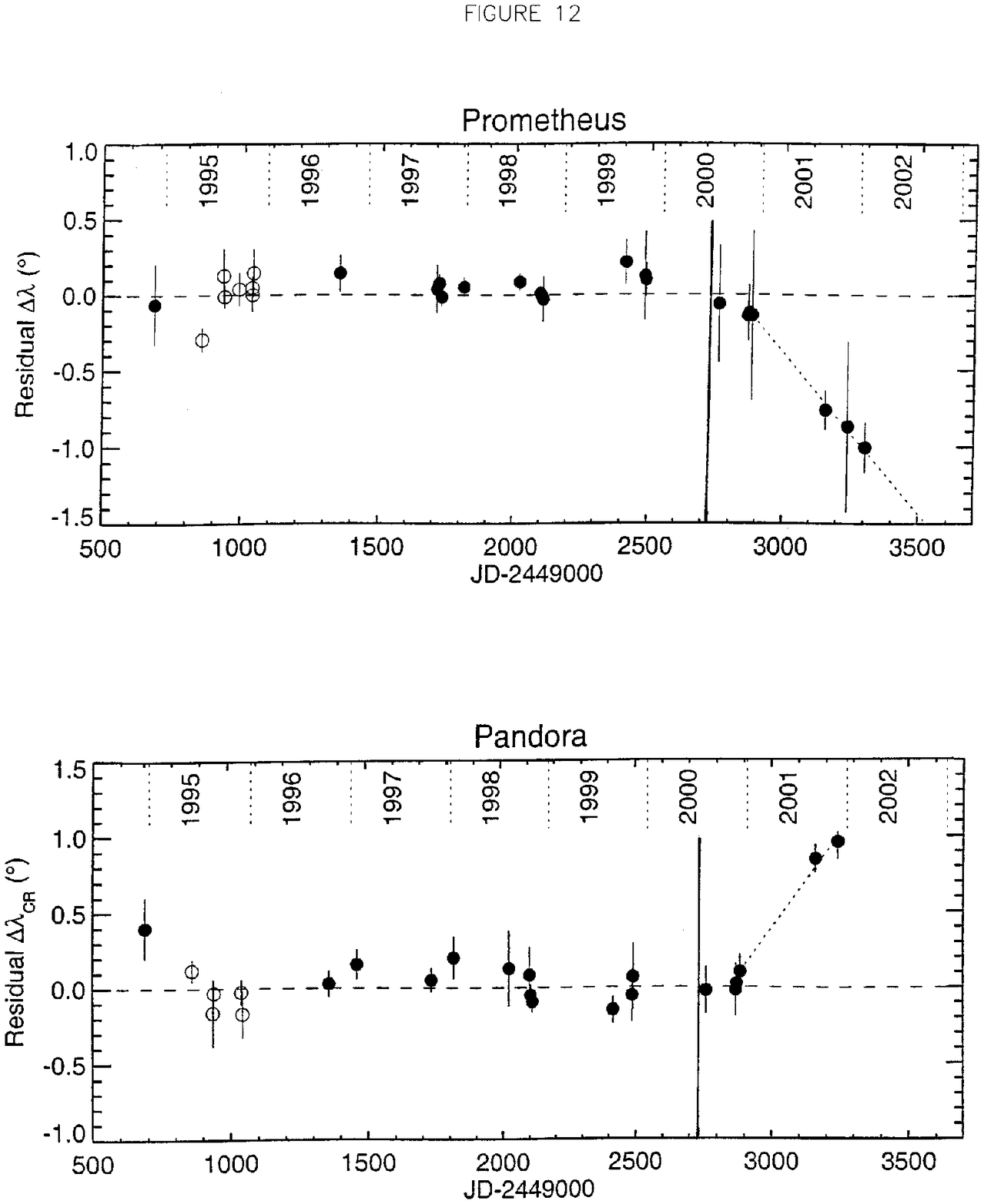}

\end{document}